\documentclass[structabstract]{aa}  
\usepackage{graphicx}
\usepackage{natbib}
\usepackage{txfonts}
\usepackage{color}
\begin{document}
   \title{Spectroscopically Identified Intermediate Age Stars
     at 0.5 - 3 pc Distance from Sgr A*}

   \authorrunning{Nishiyama et al.}
   \titlerunning{Intermediate Age Stars in the MW NSC}

   \author{Shogo Nishiyama\inst{1,2,3}, Rainer Sch\"{o}del\inst{4}, 
     Tatsuhito Yoshikawa\inst{5}, Tetsuya Nagata\inst{5}, 
     Yosuke Minowa\inst{6}, 
     and Motohide Tamura\inst{2,7}
   }

   \institute{Miyagi University of Education,
     Aoba-ku, Sendai 980-0845, Japan
     \and
     National Astronomical Observatory of Japan, 
     Mitaka, Tokyo 181-8588, Japan
     \and
     \email{shogo.nishiyama@nao.ac.jp/shogo-n@staff.miyakyo-u.ac.jp}
     \and
     Instituto de Astrof\'isica de Andaluc\'ia (CSIC),
     Glorieta de la Astronom\'ia s/n, 18008 Granada, Spain
     \and
     Department of Astronomy, Kyoto University, 
     Kyoto 606-8502, Japan
     \and
     Subaru Telescope, National Astronomical Observatory of Japan, 650 
     North A`ohoku Place, Hilo, HI 96720
     \and
     Department of Astronomy, The University of Tokyo, 
     Bunkyo-ku, Tokyo 113-0033, Japan
   }

   \date{}

 
  \abstract
  {Nuclear star clusters (NSCs) at the dynamical center of galaxies
    appear to have a complex star formation history. 
    This suggests repeated star formation
    even in the influence of the {\rm strong} tidal field from supermassive black holes.
    Although the central region of our Galaxy is an ideal target
    for studies of the star formation history in the NSCs,
    most of the past studies 
    have concentrated  on
    a projected distance of $R_{\mathrm{Sgr\,A*}} \sim 0.5$\,pc
    from the supermassive black hole Sgr\,A*.
  }
  {In our previous study,
    we have detected 31 so far unknown early-type star candidates 
    throughout the Galactic NSC
    (at $R_{\mathrm{Sgr\,A*}} = 0.5 - 3$\,pc; Nishiyama and Sch\"{o}del 2013).
    They were found via near-infrared (NIR) imaging observations
    with narrow-band filters which are sensitive to
    CO absorption lines at $\sim 2.3\,\mu$m,
    a prominent feature for old, late-type stars.
    The aim of this study is a confirmation
    of the spectral type for the early-type star candidates.
  }
  { We have carried out NIR spectroscopic observations
    of the early-type star candidates
    using Subaru/IRCS/AO188 and the laser guide star system.
    $K$-band spectra for 20 out of the 31 candidates 
    and reference late-type stars were obtained.
    By determining an equivalent width, EW(CO), 
    of the $^{12}$CO absorption feature at $\approx 2.294\,\mu$m, 
    we have derived an effective temperature and a bolometric magnitude
    for each candidate and late-type star, 
    and then constructed an HR diagram.
  }
  { No young ($\sim$ Myr), massive stars are included
    in the 20 candidates we observed;
    however, 13 candidates are most likely 
    intermediate-age giants ($50 - 500\,$Myr).
    Two other sources have ages of $\sim 1$\,Gyr,
    and the remaining five sources are old ($> 1$\,Gyr), late-type giants.
  }
  {Although none of the early-type star candidates 
    from our previous narrow-band imaging observations can be confirmed
    as a young star,
    we find that the photometric technique is sensitive to distinguish
    old, late-type giants from young and intermediate-age populations.
    In the spectroscopically observed 20 candidates, 
    65 \% of them are confirmed to be younger than $500$\,Myr.
    The intermediate-age stars could be so far unknown members of a population
    formed in a starburst $\sim 100$\,Myr ago.
    Finding no young ($\sim$ a few Myr) stars at $R_{\mathrm{Sgr\,A*}} = 0.5 - 3$\,pc
    favors the in-situ formation scenario for the presence of 
    the young stars at $R_{\mathrm{Sgr\,A*}} < 0.5$\,pc,
    although we do not exclude completely the possible existence of
     unknown young, massive stars in the region from our observations.
     Furthermore, the different spatial distributions of 
    the young and the intermediate-age stars imply that
    the Galactic NSC is an aggregate of stars
    born in different places and under different physical conditions.
  }

   \keywords{Galaxy: center -- stars: formation -- techniques: spectroscopic}
   \maketitle
%
\section{Introduction}

Most galaxies host luminous nuclear star clusters 
\citep[NSCs; e.g.,][]{Carollo98, Cote06, Boker10IAUS, Georgiev14}.
Many of them have been found to coexist with supermassive black holes (SMBHs)
at their center \citep{Seth08, Graham09, Neumayer12}.
Unlike SMBHs, 
NSCs are expected to provide a visible record of 
gas accretion and star formation at the center of galaxies;
therefore studying stellar populations in NSCs 
can give us clues as to how stars are formed 
within the strong tidal field from SMBHs.

The NSC at the center of our Galaxy is the only one
which can be resolved into individual stars with current instruments,
making it 
an ideal
target for studies of 
stellar populations and star formation history in NSCs.
A concentration of young, massive stars at the center of the Milky Way's NSC,
within a projected distance of about 0.5\,pc of the SMBH Sagittarius A* (Sgr\,A*),
was identified by spectroscopic observations
and traces a starburst that occurred a few Myr ago
\citep[e.g.,][]{Paumard06, Bartko09GC, Lu13, Yelda14}.
However, due to the strong and patchy interstellar extinction 
toward the Galactic NSC,
broad-band photometry
can hardly be used to distinguish stellar populations
\citep[see, however,][]{Schodel10Ext}.
The number of stars is too large to be surveyed 
with single or multi slit spectrographs.
The apparent size of the NSC
\citep[half-light radius of $\approx 4.2\,$pc $\approx 1\farcm8$;][]{Schodel14NSCAA}
is too large to survey the entire region of the NSC
using integral field spectrographs with a typical FoV of several arcsec,
when operating with angular resolutions on the order of 0\farcs1,
which is necessary in the crowded field of the Galactic center.
Therefore
most of the past observations of stellar populations in the Galactic NSC
have been limited to 
a region within a projected radius of $R_{\mathrm{Sgr\,A*}}\sim 0.5$\,pc from Sgr\,A*
\citep[e.g.,][]{Genzel03Cusp,Paumard06,Maness07,Do09Cusp,Bartko10,Pfuhl11,Do13,Lu13}.

In our previous paper \citep{Nishi13NSC},
we aimed at overcoming these observational limitations
by using near-infrared (NIR) imaging observations
with narrow-band filters.
CO band head absorption features starting at $2.29\,\mu$m can be used
to distinguish between massive young stars and late-type giants \citep{Buchholz09}.
The absorption features are very strong for late-type (K and M) stars,
become weaker for earlier spectral type,
and are absent for stars earlier than early-F type \citep{Wallace97ApJS}.
So we employed two narrow band filters,
2.34\,$\mu$m at, and 2.25\,$\mu$m just short-ward of, 
the CO feature,
to derive
a $[2.25] - [2.34]$\footnote{ 
[$\lambda$] denotes a magnitude in a narrow-band filter
with the central wavelength of $\lambda$.} color 
as a proxy of
stellar spectral type.
In the magnitude range of $9.75 < [2.25] < 12.25$, 
we thus found 31 so far unknown 
early-type [Wolf-Rayet (WR), supergiants or early O type] star candidates
at $R_{\mathrm{Sgr\,A*}} = 0.5 - 3$\,pc
from the SMBH Sgr\,A* (Fig. \ref{Fig:ObsSpDist}).

To determine spectral types of the early-type star candidates,
we have carried out adaptive-optics (AO) assisted
spectroscopic observations with 
the Subaru telescope\footnote{
Based on data collected at Subaru Telescope, 
which is operated by the National Astronomical Observatory of Japan.} 
and the infrared camera and spectrometer IRCS.
In this work, 
we present the results of the observations
for 20 of the above mentioned 31
early-type star candidates located outside the central 0.5 pc region.

%
\section{Observations and Data Reduction}

\begin{figure*}[htb]
  \begin{center}
  \includegraphics[width=0.9\textwidth,angle=0]{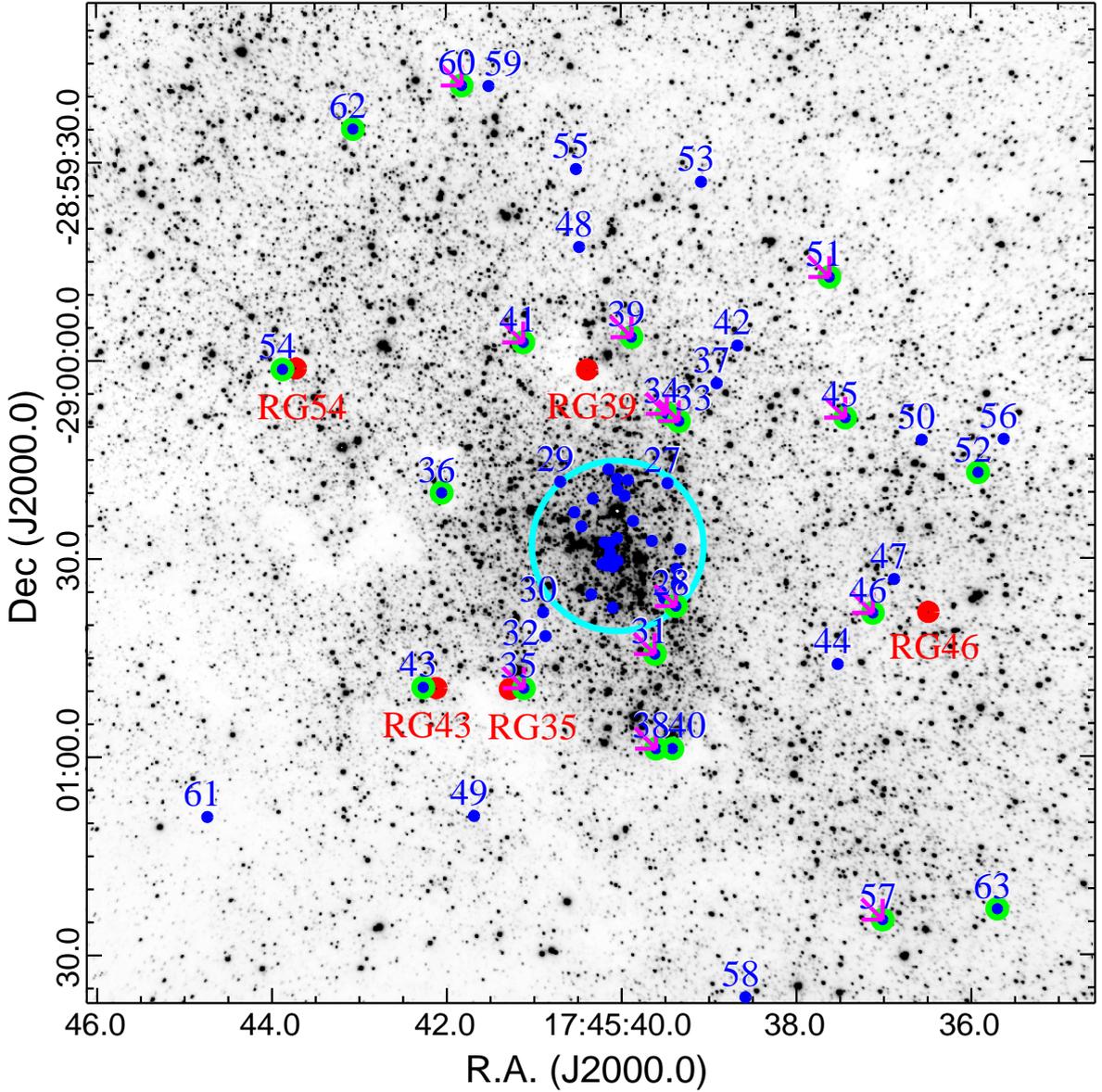}
  \caption{\label{Fig:ObsSpDist} 
    Spatial distribution of the early-type star candidates found in 
    \citet[][blue circles and ID numbers from the mentioned work]{Nishi13NSC}
    overplotted on a $2.25\,\mu$m narrow-band image
    (VLT/ISAAC).
    Spectra for 20 of the candidates were obtained with Subaru/IRCS (green circles).
    The large cyan circle delimits a region within 0.5\,pc (12\farcs9) 
    in projection from Sgr\,A*. 
    Spectra for five red giants 
    were also obtained as a reference (red circles).
    Magenta arrows represent intermediate-age (50\,Myr - 500\,Myr) stars 
    (see \S \ref{sec:SpClass} and \ref{sec:Disc}).
  }
  \end{center}
\end{figure*}

\begin{figure}[htb]
  \includegraphics[width=0.5\textwidth,angle=0]{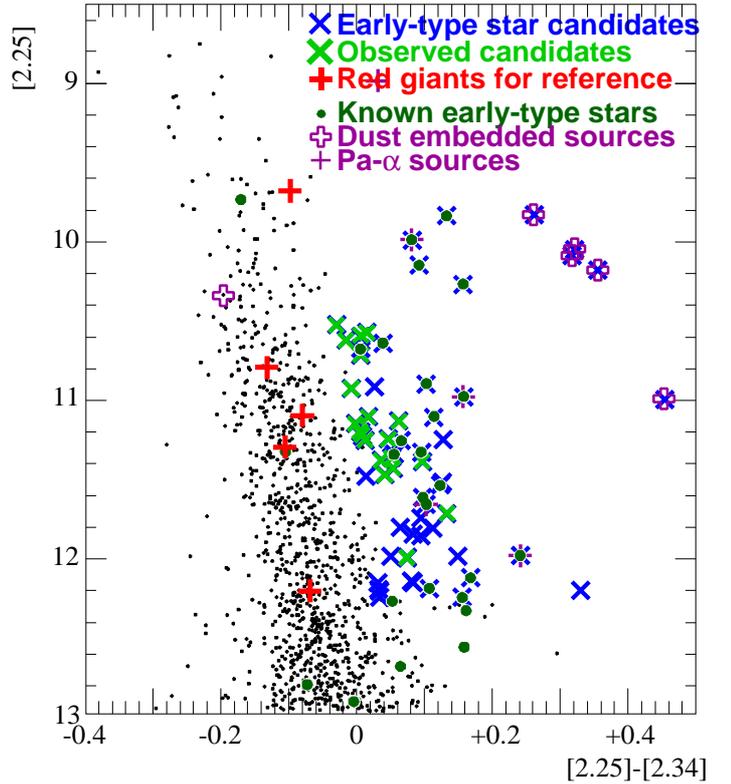}
  \caption{\label{Fig:CandsCMD} 
      [2.25] vs [2.25] - [2.34] color magnitude diagram.
      Red color (positive value) in [2.25] - [2.34] means 
      a weak CO absorption at $2.34\,\mu$m
      which is an indicator for early spectral type stars.
      A sequence of stars from $([2.25]-[2.34], [2.25])
      \sim (-0.2, 9)$ to $(0, 13)$ is the red giant branch (RGB).
      Blue ``$\times$'' are the early-type star candidates 
      found in \citet{Nishi13NSC},
      and they are distributed at the red side of the RGB,
      indicating earlier spectral type.
      Most of the bright ($[2.25] \lesssim 11$), 
      very red ($[2.25]-[2.34] \gtrsim 0.1$) are already known
      early-type stars and dust embedded sources.
      Light green ``$\times$'' represents the early-type star candidates
      whose spectrum is obtained in this study,
      and red cross are observed red giants as a spectrum reference.
  }
\end{figure}

The spectroscopic targets were selected from the early-type star candidates
found by \citet{Nishi13NSC}.
Fig. \ref{Fig:CandsCMD} shows a $[2.25]$ vs. $[2.25] - [2.34]$ color magnitude diagram
of stars in the central $2\farcm5 \times 2\farcm5$ region of our Galaxy
\citep[see also Fig. 6 in][]{Nishi13NSC}.
Red color in $[2.25] - [2.34]$ means a weak CO absorption at $\sim 2.34\,\mu$m,
and the early-type star candidates (blue ``$\times$'' in Fig. \ref{Fig:CandsCMD}) 
are distributed to the right of the red giant branch (RGB) on the color magnitude diagram.
Since most of the stars with very red $[2.25]-[2.34]$ color have already been 
identified as early-type stars (dark green circles), 
Pa-$\alpha$ sources (purple thin crosses), 
or dust embedded sources (purple thick crosses),
and there is no clear difference in the $[2.25]-[2.34]$ color among the rest of the candidates,
we have planned our observations to obtain spectra of as many targets as possible.
It means that bright stars with a smaller separation angle from Sgr\,A*
(i.e., observable without a change of natural tip-and-tilt guide stars)
have a higher priority in our observations.
As a result, we have observed 20 early-type star candidates 
(light green ``$\times$'' in Fig. \ref{Fig:CandsCMD}),
and four red giants (red crosses) as a reference.
As shown in Fig. \ref{Fig:CandsCMD}, the observed sample does not have
a strong bias on the $[2.25]-[2.34]$ color,
leading to no clear bias in the sample selection of the observed targets.

The spectroscopic observations were carried out
in the nights of 12/13 June and 4/5 August 2012
with the Subaru telescope \citep{Iye04Subaru},
and the infrared camera and spectrometer 
IRCS \citep{00KobayashiSPIE}.
The 20 observed stars are indicated by green circles in Fig. \ref{Fig:ObsSpDist}.
The exposure time was 300\,sec for all the sources except \#\,57 (150\,sec).
The IRCS grism mode provides a spectral resolution of
$\lambda/\Delta \lambda \approx 1,900$ in the $K$ band 
with a slit width of $0\farcs1$.

During our observations in August 2012, 
we used the Subaru AO system AO188 
\citep{Hayano08SPIE,Hayano10SPIE}
and the laser guide star system.
Three natural guide stars were used to correct for tip-tilt motions:
FJ95-19 (17:45:41.8, -28:59:31.0, $V=15.8$) for \#62,
FJ95-10 (17:45:39.8, -29:01:25.7, $V=15.4$) for \#57 and 63,
and USNO-A2.0 0600-28577051 (17:45:40.7, -29:00:11.2, $R=13.7$)
for the rest of our candidates.
In spite of the low elevation of the candidates at Mauna Kea,
the AO system delivers the FWHMs and Strehl ratios of
$0\farcs10 - 0\farcs17$ and $0.10 - 0.25$, respectively.
In June, since the elevation of the observed candidates
(\#28, 31, 35, 36, 38, 40) were very low,
we did not use the AO system.

The reduction process included
flat-fielding, sky subtraction, bad pixel correction, cosmic-ray removal,
wavelength calibration with an arc lamp, spectrum extraction, 
and relative flux calibration.
Flat field images were provided by obtaining spectra of continuum sources.
Interspersed with the observations, we
observed a dark cloud located at a few arcmin northwest
of the Galactic center
to obtain sky measurements.
Argon arc frames were used to fit a dispersion solution.

\begin{figure*}[!hbt]
  \includegraphics[width=0.75\textwidth,angle=0]{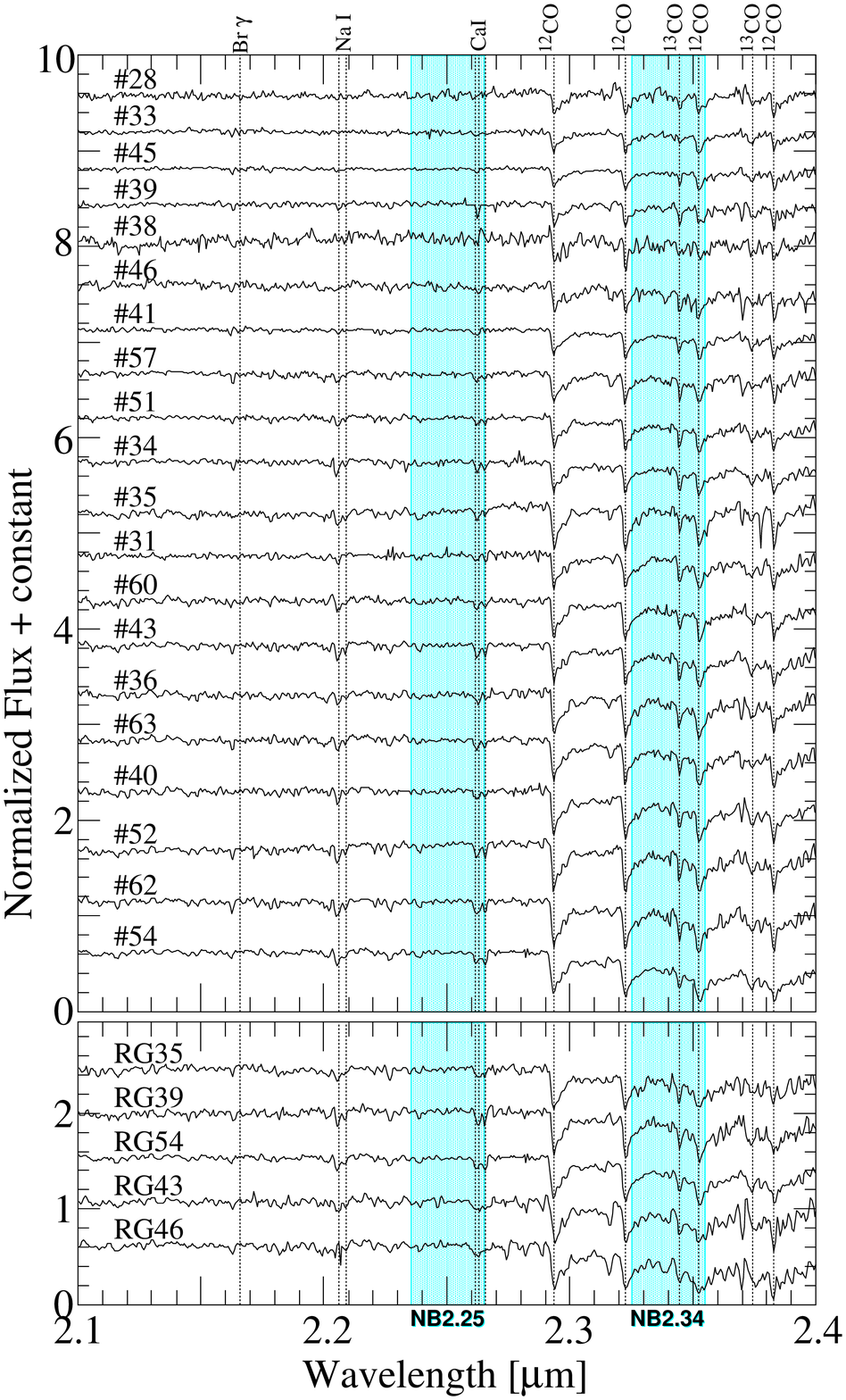}
  \caption{\label{Fig:spectra} 
    $K$-band spectra of the early-type star candidates
    (top) and RGB stars (bottom).
    Each star is identified above its spectrum
    with the ID assigned by \citet{Nishi13NSC} and Fig. \ref{Fig:ObsSpDist}.
    The position of the Br-$\gamma$, Na I doublet, Ca I triplet, 
    four band heads of $^{12}$CO,
    and two band heads of $^{13}$CO
    are indicated by the vertical broken lines.
    The position and width of two narrow-band filters,
    2.25\,$\mu$m (NB2.25) and 2.34\,$\mu$m (NB2.34), 
    are also indicated by cyan hatched boxes.
    The spectra are sorted by the strength
    of the CO absorption feature at $\approx 2.294\,\mu$m
    from the top (weak) to the bottom (strong),
    in each panel.
  }
\end{figure*}

Telluric correction was achieved by dividing each candidate's 
spectrum by that of one of the 
early-A main-sequence stars
HD 126997 (A0-1V), HD 200918 (A0V), or HD 190285 (A0V),
which were observed on the same night and at a similar airmass.
Prior to division, the Br-$\gamma$ line was removed 
from the standard star spectra by interpolating the stellar continua.
The systematic influence of the SED of the A-stars was subsequently removed 
by multiplying with
a blackbody spectrum
of an appropriate effective temperature.
Each spectrum was shifted to rest wavelength
by using 
five or six of the $^{12}$CO and $^{13}$CO bandhead absorption features
(vertical broken lines in Fig. \ref{Fig:spectra}).
We removed the curvature of the stellar continua 
by dividing the spectra by 
a third or fourth polynomial function
fitted to the line-free region of the stellar spectra.
The resultant $K$-band spectra are shown in Fig. \ref{Fig:spectra}.

As a control sample, we also observed five stars
located on the RGB
in the color-magnitude diagram 
(Fig. \ref{Fig:CandsCMD}).
The same observational settings were used
as for the early-type star candidates.

%
\section{Spectral Classification}
\label{sec:SpClass} 

$K$-band spectra provide prominent features
which can be used for stellar classification.
For early-type stars, the Br-$\gamma$ absorption line
is expected; however, no spectrum of our candidates shows 
the Br-$\gamma$ absorption.
Instead, all of the spectra show clear CO bandhead absorption features,
indicative of a spectral type later than $\sim$ G4.
So {\it none} of our 20 observed candidates is confirmed as an early-type star.

Many prominent features in the $K$-band,
such as Na I, Ca I, and CO, have been widely used
to investigate stellar parameters.
In particular the absorption strength of the $^{12}$CO$(2-0)$ bandhead 
($\approx 2.294\,\mu$m)
is a good indicator of a stellar effective temperature ($T_{\mathrm{eff}}$).
To measure the CO absorption depth, several index definitions
have been proposed \citep[see][and references therein]{08MarmolQueralto};
here we compute the
CO index according to the recipe of \citet{01Frogel}.
\citet{Pfuhl11} used this definition to determine $T_{\mathrm{eff}}$
of late-type stars in the Galactic NSC,
and have found a smaller systematic uncertainty than other definitions
used by \citet{Blum03} and \citet{Maness07}.

The definition by \citet{01Frogel} uses five narrow bandpasses,
four at continuum, and one at the CO bandhead,
to estimate the CO absorption depth \citep[see Table 2 in][]{01Frogel}.
In our study, the continuum level $\omega_{\mathrm{C}} (\lambda)$ is determined with 
a linear fit to the flux levels in the continuum bandpasses,
and the equivalent width EW(CO) is measured according to
\begin{equation}
 \mathrm{EW(CO)} = 
\int \frac{\langle \omega_{\mathrm{C}} \rangle - \omega_{\mathrm{CO}}}{\langle \omega_{\mathrm{C}} \rangle} d \lambda,
\end{equation}
where $\langle \omega_{\mathrm{C}} \rangle$ is the mean of 
continuum levels measured in the four bandpasses,
and $\omega_{\mathrm{CO}} (\lambda)$ is 
the depth of CO absorption at wavelength $\lambda$.
The uncertainty of EW(CO), $\sigma_{\mathrm{CO}}$, 
results from the uncertainty of the mean 
that is derived from the measurement in the four bandpasses.
The resultant EW(CO) and $\sigma_{\mathrm{CO}}$ for the early-type star candidates
are shown in Table \ref{Tab:StePar}.
Note that $\sigma_{\mathrm{CO}}$ only includes 
the uncertainty in the continuum level
and not any
other systematic uncertainties from, e.g., 
correction of the spectral curvature.
So the estimated $\sigma_{\mathrm{CO}}$ shown here are lower limits.
However, the expected systematic uncertainty is only 
on the order of a few percent
when we use the definition by \citet[][see also \S \ref{sec:Disc}]{01Frogel}.

For the effective temperature calibration,
we use the following EW(CO)-$T_{\mathrm{eff}}$ relation derived by \citet{Pfuhl11}:
\begin{equation}
T_{\mathrm{eff}} = 5832 - 208.28 \cdot \mathrm{EW(CO)}
+ 11.3 \cdot \mathrm{\left[EW(CO)\right]}^2  - 0.34 \cdot \mathrm{\left[EW(CO)\right]}^3 [K],
\label{eq:T-EWCO} 
\end{equation}
where $T_{\mathrm{eff}}$ is in the unit of Kelvin.
\citet{Pfuhl11} used 
the definition of EW(CO) by \citet{01Frogel},
and derived the equation above using 33 red giants
with known $T_{\mathrm{eff}}$, spectral type of G0 to M7, 
and metallicity of $-0.3 < \mathrm{[Fe/H]} < 0.2$.
This EW(CO)$- T_{\mathrm{eff}}$ relation holds in the range
$3.5 < \mathrm{EW(CO)} < 24$.
Two early-type star candidates (\#\,54 and 62) 
and a reference red giant (RG46)
have a larger EW(CO) than this range,
so that we do not determine their $T_{\mathrm{eff}}$.
It means that they are likely to be cooler than $\approx 2,700$\,K.

To determine the amount of the interstellar extinction,
we use the data sets of 1.71\,$\mu$m and 2.25\,$\mu$m 
narrow-band filters taken with VLT/ISAAC \citep{Nishi13NSC}.
A mean $[1.71]-[2.25]$ color
of the 20 nearest stars, $\langle [1.71]-[2.25] \rangle$,
was calculated at the position of each target.
Assuming that the nearest stars are late-type (K - M) giants
with an intrinsic color of $([1.71]-[2.25])_0 = 0.2$,
the amount of the interstellar extinction $A_K$ can be estimated
with the equation
$A_K \approx 1.44 ( \langle [1.71]-[2.25] \rangle - ([1.71]-[2.25])_0)$
\citep{Nishi06Ext}.
The typical uncertainty in $A_K$ determined by RMS 
of the colors the nearest stars is 0.68 mag.
Note that the intrinsic color difference of K-M giants, $\pm 0.1$,
is small enough to be safely ignored.

\begin{figure}[!htb]
  \includegraphics[width=0.5\textwidth,angle=0]{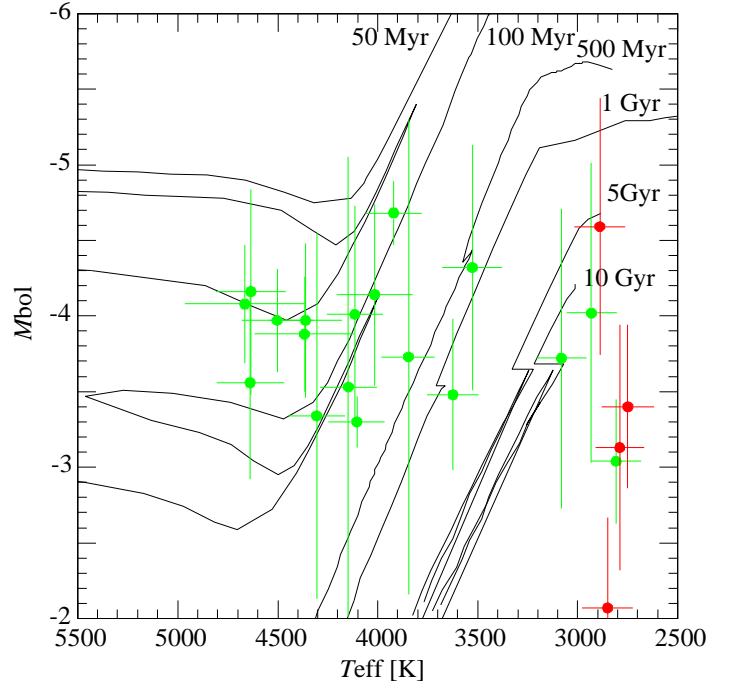}
  \caption{\label{Fig:HRD} 
    HR diagram for the early-type star candidates (green circles)
    and red giants as a reference (red circles).
    Overplotted are theoretical isochrones for ages of (from left to right)
    50\,Myr, 100\,Myr, 500\,Myr, 1\,Gyr, 5\,Gyr, and 10\,Gyr
    with solar metallicity using the Padova code \citep{Girardi00,Marigo08}.
  }
\end{figure}

To determine bolometric magnitudes,
a bolometric correction BC$_K$ is necessary.
We follow the equation 
$\mathrm{BC}_K = 2.6 - (T_{\mathrm{eff}}-3800)/1500$
from \citet{Blum03}.
With $\mathrm{BC}_K$ and $A_{K}$,
the bolometric magnitude can be calculated as
$M_{\mathrm{bol}} = K_S - A_K - \mathrm{DM} + \mathrm{BC}_K$,
where $K_S$ is the observed $K_S$-band magnitude
and DM is a distance modulus of 14.5
at the distance of 
$8.0 \pm 0.15$\,kpc \citep{Schodel10Ext}.
An uncertainty of DM of 0.04 mag is also quadratically added to the uncertainty of $M_{\mathrm{bol}}$.
Note that we ignore the negligible magnitude difference
between the $K$ and $K_S$ bands.
As a result, we obtain both $T_{\mathrm{eff}}$ and $M_{\mathrm{bol}}$,
and thus the observed stars
can be plotted on a HR diagram (Fig. \ref{Fig:HRD}).

  If supergiants are included in our targets, we need to use 
  another EW(CO)$- T_{\mathrm{eff}}$ relation for them;
  however, considering their rarity, 
  and the observed magnitudes of our targets ($K_S > 10$),
  it is safe to assume that no supergiant is included in our targets.
  In addition, the resultant bolometric magnitudes $M_{\mathrm{bol}}$ 
  of the early-type star candidates are fainter than $-4.6$ mag,
  even if we use 
  the bolometric correction of BC$_K = 2.6$ \citep{Blum03},
  and intrinsic colors of supergiants for the extinction correction;
  almost all of the supergiants
  in the central 5-pc region appear to be brighter than 
  $M_{\mathrm{bol}} \approx -5.0$ \citep{Blum03}.

\begin{figure}[!htb]
  \includegraphics[width=0.5\textwidth,angle=0]{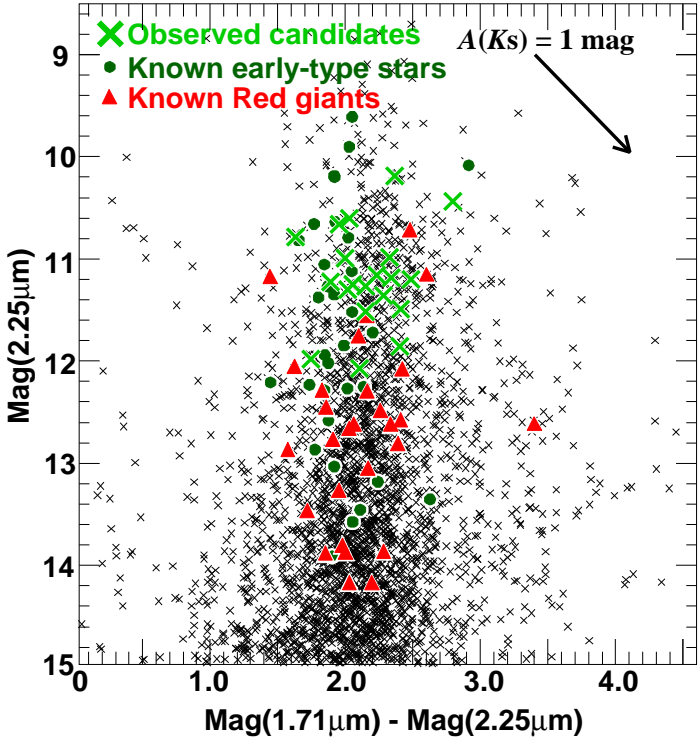}
  \caption{\label{Fig:CMD171} 
    [2.25] vs [1.71] - [2.25] color magnitude diagram
    for sources measured in \citet{Nishi13NSC}.
    The observed early-type star candidates in this work
    are overplotted by light green ``$\times$''.
    Dark green circles and red triangles
    represent spectroscopically identified early-type stars 
    and late-type giants in the Galactic NSC, respectively.
  }
\end{figure}

  Without a proper motion measurement, 
  it is still difficult to remove foreground/background sources completely.
  A less accurate but common way to remove such sources uses the interstellar extinction.
  We have constructed a narrow-band ([1.71] and [2.26]) color magnitude diagram
  for our targets, known early-type stars and red giants, and other sources
  (Fig. \ref{Fig:CMD171}).
  The $[1.71] - [2.26]$ color corresponds to NIR $H - K$, 
  and is more sensitive to the amount of the interstellar extinction 
  than $[2.25] - [2.34]$ used in Fig. \ref{Fig:CandsCMD}.
  We have confirmed that the color spread of stars observed in this work
  is almost the same as those of known early-type stars and red giants in the NSC,
  suggesting that our targets are located in the NSC as well.

%
\section{Discussion}
\label{sec:Disc} 

\subsection{Narrow-band Photometry}

The main goal of our work is a systematic search for young, massive stars 
throughout the Galactic NSC.
To find candidates of such young, massive stars in this region,
we carried out NIR imaging observations using two narrow-band filters,
$2.25\,\mu$m and $2.34\,\mu$m, with filter widths of $\approx 0.03\,\mu$m
\citep{Nishi13NSC}.
As shown in Fig. \ref{Fig:spectra}, the $2.34\,\mu$m filter is sensitive to
the CO band head absorption at $\ga 2.29\,\mu$m,
a typical feature seen in late-type stars;
on the other hand, the $2.25\,\mu$m filter is just short-ward of the CO absorption.
When we measure magnitudes of stars in the two filters, $[2.25]$ and $[2.34]$,
a redder color in $[2.25]-[2.34]$ means less CO absorption at $2.34\,\mu$m,
indicating earlier spectral type.
Hence the $[2.25]-[2.34]$ color index can be used as a proxy for EW(CO).

Unfortunately, we found strong, systematic $[2.25]-[2.34]$ color trends 
along the $x$- and $y$-axes in the ISAAC FoV \citep[Fig. 4 in][]{Nishi13NSC}, 
making absolute calibrations of $[2.25]$ and $[2.34]$ impossible.
To use the $[2.25]-[2.34]$ color to search for early-type stars, 
we carried out a color correction in the assumption that 
the average of the intrinsic stellar colors is the same throughout the observed FoV.
This assumption is appropriate for our observed field 
due to similar intrinsic colors of almost all stellar types around the $K$ band, 
the dominance of late-type giants in this region, 
and the restricted wavelength range of our observations.
Hence what we use to identify the early-type star candidates is
a {\it relative} $[2.25]-[2.34]$ color to the dominant late-type giants (RGB stars),
and the color is calibrated to have $[2.25]-[2.34] = 0$ for the RGB stars.
Then we defined stars more than 2$\sigma$ redder than 
the RGB as early-type star candidates.

\begin{figure}[!htb]
  \includegraphics[width=0.5\textwidth,angle=0]{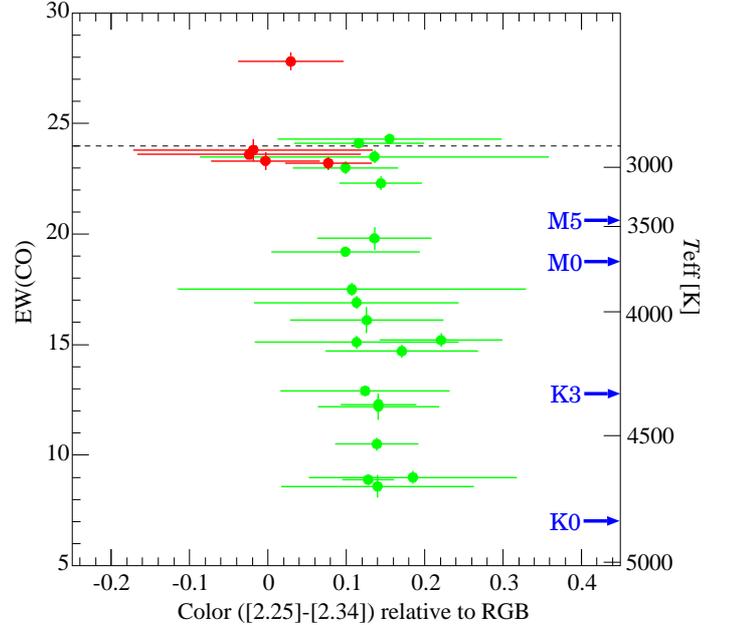}
  \caption{\label{Fig:ColCOD} 
    Relation between stellar $[2.25]-[2.34]$ color relative to
    the RGB mean color, and EW(CO)
    for the early-type star candidates (green circles) 
    and late-type giants as a reference  (red circles).
    In the relative $[2.25]-[2.34]$ color, 
    the RGB stars and early-type stars are expected to have a color of
    $[2.25]-[2.34] \approx 0$ and $[2.25]-[2.34] > 0$, respectively.
    The right-hand side axis shows the corresponding $T_{\mathrm{eff}}$
    derived by equation (\ref{eq:T-EWCO}).
    The blue arrows represents spectral type
    in the case of giants 
    \citep[according to the Table 2 in][]{Meyer98IRSpec}.
    The horizontal dashed line represents
    the upper limit of EW(CO) to determine $T_{\mathrm{eff}}$
    with equation (\ref{eq:T-EWCO}). 
  }
\end{figure}

Genuine early-type stars are expected to have 
a redder $[2.25]-[2.34]$ color than late-type giants on the RGB,
and an almost negligible EW(CO).
Fig. \ref{Fig:ColCOD} shows 
a plot of the EW(CO) of the observed stars
versus their $[2.25]-[2.34]$ color \citep[as given by][]{Nishi13NSC}.
The latter was adjusted such that 
the mean color of the RGB stars is 0
as described in the previous paragraph.
As shown by green circles in Fig. \ref{Fig:ColCOD},
all of our early-type star candidates have a color of
$[2.25]-[2.34] \gtrsim  0.1$,
and most of them have EW(CO) of $\la 20$.
This agrees well with the fact that
most of the early-type star candidates have
a smaller EW(CO) than the RGB stars.
Contrary to our expectations, however,
we do not observe any clear trend of the $[2.25]-[2.34]$ color
as a function of EW(CO).
This is probably due to relatively large uncertainties of,
and a narrow range in, the $[2.25]-[2.34]$ colors
for the early-type star candidates. 

\subsection{Results of Spectroscopic Follow-up Observations}

In order to further investigate the nature of our target stars,
we compared their position in the HR diagram (Fig. \ref{Fig:HRD})
with the theoretical isochrones for solar metallicity
\citep[][]{Girardi00,Marigo08}.
Some of our candidates are likely to be
old (age $\gtrsim 5$\,Gyr), late-type giants;
on the other hand,
most (13 out of 20)
of them are located to the left of the 500-Myr isochrone.
They are thus likely to be young giants
with a mass in the range $2.5\,M_{\sun} < M < 6\,M_{\sun}$,
and are descendants of main-sequence B-type stars.
No young ($\sim$ Myr), massive star is included
in the 20 candidates we observed.

The data points in the HR diagram might have a systematic uncertainty.
One of the candidates and 
three of the reference red giants
are located clearly to the right and below
of the oldest theoretical isochrone considering the uncertainties
shown in the plot.
\citet{Pfuhl11} investigated systematic uncertainties
in EW(CO) and $T_{\mathrm{eff}}$ due to the differences of
a spectral resolution and an amount of the interstellar extinction.
They found no measurable impact in the EW(CO) index by \citet{01Frogel}
due to degrading the resolution from $R \sim 3,000$ to $2,000$,
and the resolution in our observations is $R \sim 1,900$.
Moreover, the \citet{01Frogel} index is decreased
by only less than a few percent due to 
the uncertainty in the extinction correction
applied to the stars studied here,
leading to the combined systematic uncertainty
of less than 50\,K in $T_{\mathrm{eff}}$.

Our source \#\,33 was observed and its stellar parameters were
determined by \citet[][source ID 300]{Maness07}.
Although they used a different definition for EW(CO)
and EW(CO)-$T_{\mathrm{eff}}$ relation,
their resultant $T_{\mathrm{eff}}$ and $M_{\mathrm{bol}}$ are
4529\,K and $-3.58$, respectively;
both of them are in very good agreement with ours,
$T_{\mathrm{eff}} = 4638 \pm 169$ 
and $M_{\mathrm{bol}} = -3.6 \pm 0.2$.
This suggests no strong systematic offset in our determination
of $T_{\mathrm{eff}}$ and $M_{\mathrm{bol}}$,
and that we have successfully found an intermediate-age population 
in the Galactic NSC.
In any case,
we do not think that the possible systematic uncertainties
present in our HR diagram affect its main features;
we can clearly distinguish two populations,
an older, cooler one and a younger hotter one.

\citet{Pfuhl11} found a significant population of outliers
which are distributed to the right and below of the oldest isochrone for solar metallicity.
They ascribed the presence of the outliers to possible effects 
such as dust envelopes, variability, high metallicity, 
or uncertainty in stellar evolutionary models.

Several outliers, distributed to the right and below of the 10-Gyr isochrone,
are also found in our HR diagram (Fig. \ref{Fig:HRD}).
We have made a conservative estimate of the uncertainties in $A_K$ (\S 3),
leading to a typical $\sigma_{{M_{\mathrm{bol}}}}$ of 0.7 mag.
Fig. \ref{Fig:CMD171} clearly demonstrates that the candidates
must be located very close to the Galactic center;
even if they are located at 1\,kpc from the Galactic center,
$M_{\mathrm{bol}}$ would be changed only 0.3 mag,
and they are thus still outliers.

\begin{figure}[!htb]
  \includegraphics[width=0.5\textwidth,angle=0]{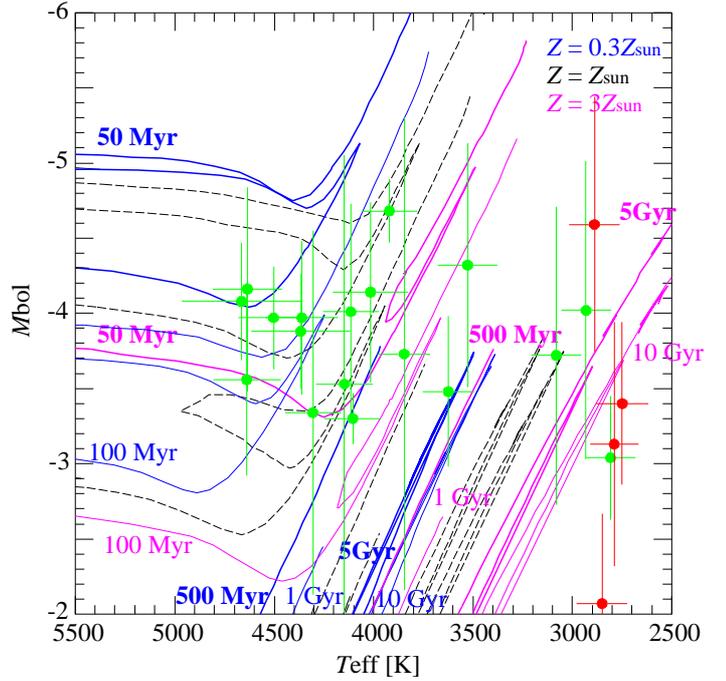}
  \caption{\label{Fig:HRDmetal} 
    HR diagram as same as Fig. \ref{Fig:HRD},
    but overplotted are theoretical isochrones for ages of (from left to right)
    50\,Myr, 100\,Myr, 500\,Myr, 1\,Gyr, 5\,Gyr, and 10\,Gyr
    with $Z = 0.3Z_{\sun}$ (blue lines), $Z = Z_{\sun}$ (black dotted lines), 
    and $Z = 3Z_{\sun}$ (magenta lines),
    using the PARSEC isochrones \citep{Bressan12PARSEC}.
    Note that the PARSEC isochrones do not include the tip of the AGB,
    but it can be used to compare isochrones for different metallicities.
  }
\end{figure}

The outliers can be potentially explained by their super-solar metallicity.
In Fig. \ref{Fig:HRDmetal}, 
we have constructed an HR diagram with isochrones of different metallicities.
Here the PARSEC isochrones \citep{Bressan12PARSEC} with 
$Z = 0.3Z_{\sun}$ (blue lines), $Z_{\sun}$ (black dotted lines), 
and $3Z_{\sun}$ (magenta lines) are used.
The diagram implies that most of the outliers are
very old ($\sim 10\,$Gyr) population with $Z \ga 3Z_{\sun}$,
and this has already been pointed out by \citet{Pfuhl11}.
Observational constraints of the metallicity of the outliers
are crucial to understand their true nature.

Stars in dense stellar cusps around SMBHs suffer much more close tidal encounters
than in normal environments,
and tidal spin-up of stars are expected in the Galactic center region \citep{Alexander01}.
The rotation of stars has an impact on the stellar evolution, and 
isochrones from \citet{Ekstrom12} are thus used
to understand the effect of the rotation. 
When the stellar rotation is considered,
isochrones are shifted towards a higher luminosity.
This cannot explain the existence of the outliers
distributed to the right and below of the oldest isochrone
in Fig. \ref{Fig:HRD}.

In the previous paper, we concluded that we found 
strong candidates for early-type stars \citep{Nishi13NSC}.
However, as shown in Fig. \ref{Fig:spectra} and \ref{Fig:HRD},
{\it no} young, massive star is included in our observed candidates;
instead, most of them are an intermediate-age (50\,Myr - 500\,Myr) population.
We note that precision-photometry in our previous study was limited to
the magnitude range of $9.75 < [2.25] < 12.25$.
Only WR stars, supergiants and early-O type stars are distributed in this range as early-type stars,
and their expected number is rather small.
B-type stars, which may be more frequent,
have magnitudes around $K \sim 14 - 16$ in the Galactic center.
They could not have been found in our seeing-limited ISAAC observations,
where the extreme crowding in the NSC 
severely limits completeness and photometric precision.

  In our previous paper, 
  we have shown an azimuthally averaged, projected stellar surface density plot
  as a function of the distance to Sgr\,A* for the early-type star candidates
  \citep[Fig. 11 in][]{Nishi13NSC}.
  It shows a continuous profile in the range from 1\farcs5 to 60\arcsec
  with a power-law index of 1.60.
  However, again, {\it no} young, massive star is included in our observed candidates.
  It means that the profile we have made in the previous paper
  is not for the early-type, young and massive stars,
  but for stars younger than $\sim 500$ Myr.
  Our results also indicate a lack of young, massive stars 
  outside the central 0.5\,pc region from Sgr\,A*;
  recently, \citet{Stostad15} have found a break 
  in the surface density profile of young stars ($\sim 5\,$Myr)
  at 0.52\,pc ($\sim 13\arcsec$) from Sgr\,A*.
  This is consistent with the non-detection of genuine young, massive stars
  in our spectroscopic follow-up observations.

Although we have not found new young, massive (WR, supergiants or early O type) stars,
it does not mean that there are no unknown young stars 
in the $R_{\mathrm{Sgr\,A*}} = 0.5 - 3$\,pc region.
As shown in Fig. \ref{Fig:CandsCMD}, 
we have misidentified a few, already known early-type stars as RGB stars.
In addition, we have not completed spectroscopic observations 
for the early-type star candidates we have found in the previous work.
Therefore we do {\it not} exclude the possibility of the existence of 
unknown early-type stars at $0.5 - 3$\,pc from the SMBH.

The detection of the intermediate-age population shows that
although the technique to measure the CO absorption depth 
with narrow-band filters cannot distinguish
between young and intermediate-age stars,
it works well to distinguish them from old, late-type giants.
In our previous paper, we have estimated that
the contamination rate of the early-type candidates by erroneous identification
of late-type giants is about 20 \% \citep{Nishi13NSC}.
As shown in Fig. \ref{Fig:ColCOD},  15 out of the 20 candidates show
$T_{\mathrm{eff}} > 3,500\,K$, and they are clearly hotter than 
the late-type giants with $T_{\mathrm{eff}} \la 3,000\,K$,
leading to a contamination rate of $25$\,\%.
However, we note that we cannot distinguish the young ($\sim$ Myr), massive stars
from the intermediate-age ($< 500$\,Myr) stars in our imaging observations.
As shown in Fig. \ref{Fig:CandsCMD}, no clear difference is seen in the distributions
of the spectroscopically confirmed massive, early-type stars and the intermediate-age 
population. Hence spectroscopic follow-up observation is necessary to discriminate
genuine young, massive stars from the sample selected by 
our imaging observations with two narrow-band filters.

\begin{figure}[!htb]
  \includegraphics[width=0.5\textwidth,angle=0]{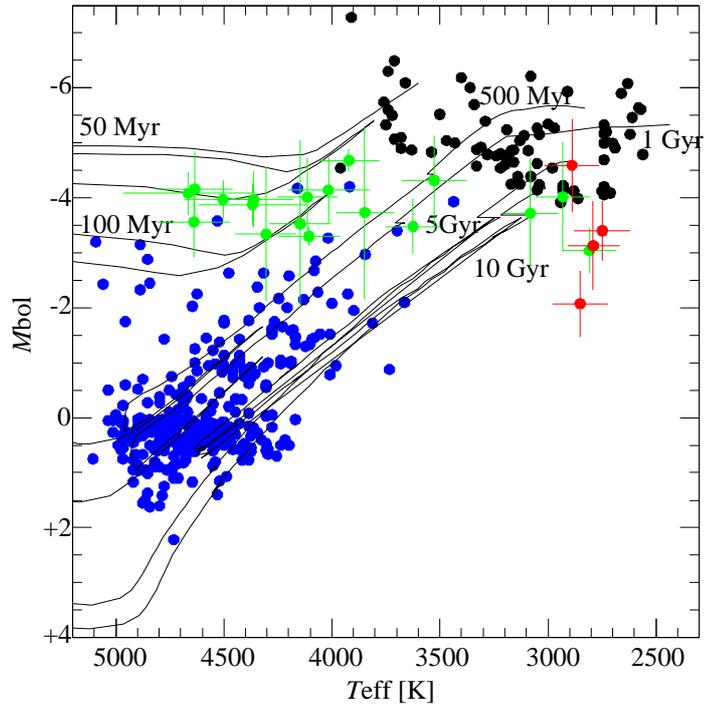}
  \caption{\label{Fig:HRDall} 
    HR diagram as same as Fig. \ref{Fig:HRD},
    but stars observed by \citet{Blum03} and \citet{Maness07}
    are also plotted with black and blue circles, respectively.
    Overplotted are theoretical isochrones for ages (from left to right)
    50\,Myr, 100\,Myr, 500\,Myr, 1\,Gyr, 5\,Gyr, and 10\,Gyr
    with solar metallicity using the Padova code \citep{Girardi00,Marigo08}.
  }
\end{figure}

We present an HR diagram
including results of previous works \citep{Blum03, Maness07}
in Fig. \ref{Fig:HRDall}. 
Since the observations by \citet{Maness07} were
assisted by AO, their 50\,\% completeness limit is as deep as
$K_S \sim 15.5$,  about 4-mag deeper than ours;
but note that the observations of \citet{Maness07} were limited
to eight relatively small fields
($4\farcs2 \times 4\farcs2$ or $0.16 \times 0.16$\,pc)
within a projected distance of $R_{\mathrm{Sgr\,A*}} = 1$\,pc from Sgr\,A*.
On the other hand, the completeness of \citet{Blum03},
a magnitude-limited survey 
that encompassed the entire region of our ISAAC survey
with a 4-m class telescope,
appears to be slightly shallower than ours.
This can probably be attributed to
the excellent seeing during our ISAAC observations.
Our findings also indicate that observations 
with higher angular resolutions, e.g., with AO,
will allow us to reach
a deeper limiting magnitude and 
search for late-O and B-type stars
via narrow-band imaging.

\subsection{Implications for Star Formation at the Galactic Center}

Finding no WR, early-O main sequence and O supergiant stars 
at the $R_{\mathrm{Sgr\,A*}}= 0.5 - 3\,$pc region
disfavors the cluster infalling scenario 
for the presence of young stars in the central 0.5\,pc region
\citep[e.g.,][]{Gerhard01,Kim03}.
This scenario proposed
the formation of a massive cluster at more than several parsec distance from Sgr\,A*,
where the tidal field from the SMBH is weak enough to form stars,
followed by an infall of the cluster toward the central parsec. 
Since the star cluster migrates to the center via dynamical friction,
a presence of some very massive stars at $R_{\mathrm{Sgr\,A*}} > 0.5$\,pc,
which escaped from the cluster is expected \citep[e.g.,][]{Fujii10GC},
but no such stars have been found in our observations,
and only two young, massive star candidates have been found
in the recent HST Pa\,$\alpha$ survey
[\# 38 and \# 133 in the list of \citet{Dong11HST},
but \# 38 is likely to be a foreground O4-6I star \citep[][star \# 7]{Mauerhan10IRW}].
Considering the sensitivity for the Pa\,$\alpha$ emission
(from evolved massive stars with strong wind) of the HST survey,
and that for the CO absorption in our observations,
the results described above
provide further evidence to support the in-situ formation scenario.

However, we emphasize that we cannot exclude 
the cluster infall scenario in general.
A migration of stellar clusters to the center is a natural consequence
if clusters are formed in the Galactic center region.
Massive stars tend to be carried very close to the center after migration,
because massive stars sink to the cluster center due to the mass segregation,
and stars at the outer region of the cluster easily become unbound \citep{Fujii10GC}.
The lifetime of the intermediate-age population stars we have found,
more than 50\,Myr, is long enough for clusters to migrate
from a few tens parsec distance 
to the central a few parsec region \citep[e.g.,][]{Gurkan05GC}.
Hence the intermediate-age population we have found
may be due to either in-situ formation or cluster infall.
In the latter case, measuring the kinematics of 
the intermediate-age stars may give some clue as to their origin,
given that the two-body relaxation time at their location is expected
to be at least an order of magnitude longer than their lifetimes
\citep{Alexander05PhR}.

A large part of our intermediate-age population
might have been formed in a starburst about 100\,Myr ago.
From the detection of a fairly large number ($\sim 10$) of
moderately luminous late-type stars,
a starburst of 100\,Myr ago was suggested 
within a projected distance of 0.5\,pc of Sgr\,A* \citep{Krabbe95}.
Further spectroscopic observations confirmed
the presence of such stars
inside \citep{Maness07} and outside \citep{Blum03} of the central 0.5\,pc region. 
The best-fit model for the star formation history 
of the Galactic NSC by \citet[][their Fig. 14]{Pfuhl11} suggests that
the star formation rate reached a minimum $\sim 1\,$Gyr ago,
and then rose again $\sim 100\,$Myr ago.
Hence the intermediate-age stars discovered by us may represent
so far unknown members of the population formed $\sim 100$\,Myr ago.

The different spatial distribution
between the young (a few Myr) and intermediate-age population stars is intriguing.
Most of the young stars are concentrated 
in the central 0.5\,pc region 
\citep[e.g.,][]{Eckart95ApJ, Genzel00MNRAS, Paumard06, Do13},
while the intermediate-age stars are distributed throughout the NSC
\citep{Haller89,Blum96AJ,Blum03}.
Our observations further confirmed the widespread distribution 
of the intermediate-age stars compared to the young massive stars.
It is unlikely that the widespread distribution 
results from dynamical scattering of stars
formed in the central 0.5\,pc,
because the intermediate-age stars are relatively heavy ($M \ga 2.5\,M_{\sun}$).
The two-body relaxation and mass segregation time scales in the NSC
are on the order of $10^9$\,yr \citep{Alexander05PhR},
i.e., we can expect that the kinematics and distribution of
the intermediate-age stars still bears the fingerprint of their origin.
As concerns the origin of these young stars,
it is not difficult for star clusters 
to reach the central a few parsec region from a few tens parsec distance
within $\sim 100$\,Myr via dynamical friction.
Alternatively the stars might have been formed in the circumnuclear disk \citep{YusefZadeh08CND}. 
In any case, our results imply
several paths of the star formation in the Galactic NSC,
i.e., stars in the NSC have been formed in different places
and under different physical conditions.

%
\section{Summary}
\label{sec:Summary} 

We have carried out spectroscopic observations of 
20 out of 31 early-type star candidates
in the nuclear star cluster at the center of our Galaxy,
identified in the imaging survey by \citet{Nishi13NSC}.
We have found that 65 \% of the candidates 
probably belong to
an intermediate-age (50\,Myr - 500\,Myr) population,
and the rest of them are late-type giants older than $\sim 1$\,Gyr.
The intermediate-age population stars are likely to have formed
in a starburst about 100\,Myr ago.
None of the stars is as young as the few million year old stars
within a projected radius of $R_{\mathrm{Sgr\,A*}} = 0.5$\,pc of Sgr\,A*.
We can thus conclude that the most recent star formation episode
in the nuclear star cluster
was confined to the immediate environment of Sgr\,A*,
at least as concerns the most massive stars.
Our findings thus also supports the in-situ formation 
of the O/WR stars near Sgr\,A*
and speaks against the cluster-infall scenario.
In the latter case, we would have expected
to observe at least a few O/WR stars in the region beyond $R_{\mathrm{Sgr\,A*}}=0.5\,$pc.
We show that narrow-band imaging observations
can be an efficient means to distinguish 
younger ($< 500$\,Myr) stars from very old giants;
however, our results also suggest that 
the kind of seeing-limited photometry with only two filters 
that we used in our previous work
is not sufficient to discriminate
young ($< 10$\,Myr), massive stars from 
the intermediate-age ($10 - 500$\,Myr) ones.
Spectroscopic follow-up observations still play an important roll
in the separation.
The clearly different distribution of the young and intermediate-age stars 
in the Milky Way's nuclear star cluster may indicate
different formation scenarios for the two populations.

\begin{acknowledgements}
  This work was supported by KAKENHI, Grant-in-Aid for 
  Research Activity Start-up 23840044, 
  Specially Promoted Research 22000005,
  COE Research 23103001 and 24103508,
  Grant-in-Aid for Exploratory Research 15K13463,
  Grant-in-Aid for challenging Exploratory Research 15K13463,
  and Young Scientists (A) 25707012,
  and Institutional Program for Young Researcher Overseas Visits.  
  RS acknowledges support by grants AYA2010-17631 and AYA2009-13036 
  of the Spanish Ministry of Economy and Competition,
  and by grant P08-TIC-4075 of the Junta de Andaluc\'ia. 
  RS acknowledges support by the Ram\'on y Cajal programme of 
  the Spanish-Ministry of Economy and Competition.
  This material is partly based upon work supported in
  part by the National Science Foundation Grant No. 1066293 
  and the hospitality of the Aspen Center for Physics.
  The research leading to these results has received funding 
  from the European Research Council 
  under the European Union's Seventh Framework Programme 
  (FP/2007-2013)/ERC Grant Agreement No. [614922].
\end{acknowledgements}

%
\begin{table*}[htb]
\caption{Observed candidates and their parameters. \label{Tab:StePar}} 
\begin{tabular}{ccccccccccccc}
\hline
\hline
ID\tablefootmark{a} & RA\tablefootmark{b} & Dec\tablefootmark{b} 
& $R_{\mathrm{Sgr\,A*}}$\tablefootmark{c} & $H$\tablefootmark{d} 
& $K_S$\tablefootmark{d} & $A_{K}$
& EW(CO) & $\sigma_{\mathrm{CO}}$ & $T_{\mathrm{eff}}$ & $\sigma_{T_{\mathrm{eff}}}$ 
& $M_{\mathrm{bol}}$ & $\sigma_{{M_{\mathrm{bol}}}}$ \\
& (J200.0) & (J200.0) & [\arcsec] & [mag] & [mag] & [mag] 
& [\AA] & [\AA] & [K] & [K] & [mag] & [mag]  \\
\hline
28 & 17:45:39.375 & $-29:00:37.21$ & 12.61 & 13.3 & 11.0 & 2.6 &   8.6 & 0.5 & 4665 & 299 & $-4.1$ & 0.4 \\      
31 & 17:45:39.616 & $-29:00:44.46$ & 17.28 & 12.1 & 10.2 & 2.9 & 16.9 & 0.3 & 3921 & 142 & $-4.7$ & 0.2 \\      
33 & 17:45:39.339 & $-29:00:09.22$ & 21.00 & 14.0 & 11.8 & 2.9 &   8.9 & 0.2 & 4638 & 169 & $-3.6$ & 0.6 \\      
34 & 17:45:39.475 & $-29:00:08.12$ & 21.31 & 14.3 & 11.5 & 2.7 & 15.2 & 0.3 & 4105 & 141 & $-3.3$ & 0.2 \\      
35 & 17:45:41.114 & $-29:00:49.62$ & 25.72 & 13.6 & 10.9 & 3.0 & 16.1 & 0.6 & 4015 & 190 & $-4.1$ & 0.6 \\      
36 & 17:45:42.052 & $-29:00:20.02$ & 27.60 & 13.4 & 10.6 & 3.2 & 19.8 & 0.5 & 3527 & 150 & $-4.3$ & 0.8 \\      
38 & 17:45:39.603 & $-29:00:58.80$ & 31.23 & 13.7 & 11.7 & 3.3 & 12.2 & 0.6 & 4366 & 250 & $-3.9$ & 0.4 \\      
39 & 17:45:39.880 & $-28:59:56.49$ & 31.68 & 13.2 & 11.2 & 2.8 & 10.5 & 0.3 & 4503 & 179 & $-4.0$ & 0.3 \\      
40 & 17:45:39.416 & $-29:00:58.78$ & 31.75 & 13.9 & 10.7 & 3.4 & 23.0 & 0.3 & 2931 & 126 & $-4.0$ & 1.0 \\      
41 & 17:45:41.117 & $-28:59:57.27$ & 33.91 & 13.6 & 11.4 & 2.5 & 12.9 & 0.2 & 4305 & 141 & $-3.3$ & 1.2 \\      
43 & 17:45:42.266 & $-29:00:49.47$ & 36.19 & 14.0 & 11.4 & 3.1 & 19.2 & 0.2 & 3624 & 127 & $-3.5$ & 0.5 \\      
45 & 17:45:37.436 & $-29:00:08.73$ & 39.27 & 13.5 & 11.3 & 3.0 &   9.0 & 0.3 & 4634 & 176 & $-4.2$ & 0.7 \\      
46 & 17:45:37.122 & $-29:00:38.29$ & 39.61 & 13.8 & 11.5 & 3.2 & 12.3 & 0.4 & 4361 & 184 & $-4.0$ & 0.5 \\      
51 & 17:45:37.616 & $-28:59:47.42$ & 51.63 & 13.6 & 11.1 & 3.0 & 15.1 & 0.3 & 4114 & 139 & $-4.0$ & 0.7 \\      
52 & 17:45:35.917 & $-29:00:17.02$ & 55.21 & 12.7 & 10.7 & 2.5 & 23.5 & 0.3 & 2809 & 126 & $-3.0$ & 0.4 \\      
54 & 17:45:43.872 & $-29:00:01.32$ & 56.96 & 13.4 & 11.2 & 2.7 & 24.3 & 0.2 & $\la$2700 & -- & -- & --\\      
57 & 17:45:37.013 & $-29:01:24.71$ & 69.15 & 13.9 & 11.2 & 2.6 & 14.7 & 0.3 & 4146 & 143 & $-3.5$ & 1.5\\      
60 & 17:45:41.819 & $-28:59:18.38$ & 73.52 & 13.4 & 11.0 & 2.8 & 17.5 & 0.3 & 3847 & 133 & $-3.7$ & 1.6\\      
62 & 17:45:43.062 & $-28:59:24.95$ & 74.56 & 14.0 & 11.3 & 2.9 & 24.1 & 0.2 & $\la$2700 &  -- & -- & --\\      
63 & 17:45:35.700 & $-29:01:23.08$ & 79.14 & 12.4 & 10.3 & 2.6 & 22.3 & 0.3 & 3081 & 125 & $-3.7$ & 1.0\\      
\hline
\end{tabular}
\tablefoot{
\tablefoottext{a}{ID from Table 2 in \citet{Nishi13NSC}.}
\tablefoottext{b}{Typical positional uncertainty is $0\farcs1 - 0\farcs3$.}
\tablefoottext{c}{Distance from Sgr\,A*.}
\tablefoottext{d}{From \citet{Nishi06Ext}.}
}
\end{table*}

\begin{table*}[htb]
\caption{Parameters for reference red giants. \label{Tab:SteParRG}} 
\begin{tabular}{cccccccccccc}
\hline
\hline
ID\tablefootmark{a} & RA\tablefootmark{b} & Dec\tablefootmark{b} 
& $H$\tablefootmark{c} & $K_S$\tablefootmark{c} & $A_{K}$
& EW(CO) & $\sigma_{\mathrm{CO}}$ & $T_{\mathrm{eff}}$ & $\sigma_{T_{\mathrm{eff}}}$ 
& $M_{\mathrm{bol}}$ & $\sigma_{M_{\mathrm{bol}}}$\\
& (J200.0) & (J200.0) & [mag] & [mag] & [mag] 
& [\AA] & [\AA] & [K] & [K] & [mag] & [mag] \\
\hline
RG35 & 17:45:41.273 & $-29:00:49.73$ & 12.7 &   9.9 & 3.2 & 23.2 & 0.3 & 2889 & 127 & $-4.6$ & 0.8\\
RG39 & 17:45:40.389 & $-29:00:01.41$ &   --   & 12.4 & 3.2 & 23.3 & 0.4 & 2851 & 127 & $-2.1$ & 0.6\\
RG43 & 17:45:42.117 & $-29:00:49.58$ & 13.9 & 11.2 & 3.4 & 23.8 & 0.5 & 2749 & 130 & $-3.4$ & 0.5 \\
RG46 & 17:45:36.488 & $-29:00:38.14$ & 14.0 & 11.2 & 3.2 & 27.8 & 0.4 &  $\la$2700 &  -- &     --  &  -- \\
RG54 & 17:45:43.721 & $-29:00:01.17$ & 13.5 & 11.1 & 3.0 & 23.6 & 0.2 & 2790 & 122 & $-3.1$ & 0.8\\
\hline
\end{tabular}
\tablefoot{
\tablefoottext{a}{See Fig. \ref{Fig:ObsSpDist}.}
\tablefoottext{b}{Typical positional uncertainty is $0\farcs1 - 0\farcs3$.}
\tablefoottext{c}{From \citet{Nishi06Ext}.}
}
\end{table*}



\end{document}